\documentstyle[amssymb,floats,aps,prb,epsf]{revtex}
\epsfclipon
\begin{document}
\twocolumn[\hsize\textwidth\columnwidth\hsize\csname
@twocolumnfalse\endcsname

\draft \draft
\title{Doping dependence of charge dynamics in electron-doped
cuprates}
\author{Tianxing Ma, Huaiming Guo, and Shiping Feng}
\address{Department of Physics, Beijing Normal University, Beijing
100875, China}

\maketitle
\begin{abstract}
Within the $t$-$t'$-$J$ model, the doping dependence of charge
dynamics in the electron-doped cuprates is studied. The
conductivity spectrum shows an unusual pseudogap structure with a
low-energy peak appearing at $\omega\sim 0$ and an rather sharp
midinfrared peak appearing around $\omega\sim 0.3\mid t\mid$, and
the resistivity exhibits a crossover from the high temperature
metallic-like to low temperature insulating-like behavior in the
relatively low doped regime, and a metallic-like behavior in the
relatively high doped regime, in qualitative agreement with
experiments. Our results also show that the unusual pseudogap
structure is intriguingly related to the strong antiferromagnetic
correlation in the system.
\end{abstract}

]
\bigskip

\narrowtext

The parent compounds of cuprate superconductors are believed to
belong to a class of materials known as Mott insulators with the
antiferromagnetic (AF) long-range order (AFLRO), then
superconductivity occurs by electron or hole doping
\cite{kastner,anderson}. It has been found from experiments that
only an approximate symmetry in the phase diagram exists about the
zero doping line between the electron- and hole-doped cuprates
\cite{alff}, and the significantly different behavior of the
electron- and hole-doped cuprates is observed
\cite{tokura,shen,yamada}, reflecting the electron-hole asymmetry.
For the hole-doped cuprates \cite{kastner}, AFLRO disappears
rapidly with doping, and is replaced by a disordered spin liquid
phase with characteristics of incommensurate short AF
correlations, then the systems become superconducting over a wide
range of the hole doping concentration $\delta$, around the
optimal $\delta\sim 0.15$. However, AFLRO survives until
superconductivity appears over a narrow range of $\delta$ around
the optimal $\delta\sim 0.15$ in the electron-doped cuprates,
where the underdoped superconducting regime is absent, and the
maximum achievable superconducting transition temperature is much
lower than the hole-doped cuprates \cite{alff,tokura1}. Moreover,
commensurate spin response in the superconducting state of the
electron-doped cuprates is observed \cite{yamada}. Therefore the
investigating similarities and differences of the hole- and
electron-doped cuprates would be crucial to understanding physics
of the high-T$_{c}$ superconductivity.

Among the striking features of the normal-state properties, the
physical quantity which most evidently displays the
nonconventional property is the charge dynamics
\cite{wang,zhang,homes,onose}, which is manifested by the optical
conductivity and resistivity. Although the resistivity shows a
tendency to deviate from the linear behavior in temperature for
the hole-doped cuprates \cite{zhang,onose}, a pseudogaplike
feature in the electron-doped cuprates has been observed in the
optical conductivity \cite{homes,onose}. This pseudogap energy
increases with decreasing doping \cite{onose,sawa}. Moreover, a
clear departure from the universal Wiedemann-Franz law for the
typical Fermi-liquid behavior is observed \cite{hill}, which shows
that the anomalous charge transport of the electron-doped cuprates
does not fit in the conventional Fermi-liquid theory. These
unusual physical properties of the charge dynamics in the
electron-doped cuprates offer the unique possibility of comparison
with those in their hole-doped counterparts which can serve as a
touchstone for theory. We \cite{feng1} have developed a
fermion-spin theory based on the charge-spin separation to study
the physical properties of the hole-doped cuprates, where the
electron operator is decoupled as the gauge invariant dressed
holon and spinon. Within this theory, we have discussed the doping
dependence of the charge transport in the hole-doped cuprates
\cite{feng1,feng2}, and the results are in agreement with
experiments \cite{kastner}. It has been shown that the charge
transport of the hole-doped cuprates is mainly governed by the
scattering from the dressed holons due to the dressed spinon
fluctuation \cite{feng1,feng2}. In this paper, we apply this
successful approach to study the doping dependence of the charge
dynamics in the electron-doped cuprates. Within the $t$-$t'$-$J$
model, we show that as in the hole-doped cuprates the unusual
behavior of the charge dynamics in the electron-doped cuprates is
intriguingly related to the AF correlations.

In the electron-doped cuprates, the characteristic feature is the
presence of the two-dimensional CuO$_{2}$ plane
\cite{tokura,shen,yamada} as in the hole-doped case, and it seems
evident that the unusual behaviors are dominated by this plane. It
is believed that the essential physics of the electron-doped
CuO$_{2}$ plane is contained in the $t$-$t'$-$J$ model on a square
lattice \cite{anderson},
\begin{eqnarray}
H&=&t\sum_{i\hat{\eta}\sigma}PC_{i\sigma}^{\dagger}
C_{i+\hat{\eta}\sigma}P^{\dagger}-t'\sum_{i\hat{\tau}\sigma}P
C_{i\sigma}^{\dagger}C_{i+\hat{\tau}\sigma}P^{\dagger} \nonumber \\
&-&\mu\sum_{i\sigma}P C_{i\sigma}^{\dagger}C_{i\sigma}P^{\dagger}
+J\sum_{i\hat{\eta}} {\bf S}_{i} \cdot{\bf S}_{i+\hat{\eta}},
\end{eqnarray}
where $t<0$ and $t'<0$ in the electron-doped case, $\hat{\eta}
=\pm\hat{x},\pm\hat{y}$, $\hat{\tau}=\pm\hat{x}\pm\hat{y}$,
$C^{\dagger}_{i\sigma}$ ($C_{i\sigma}$) is the electron creation
(annihilation) operator, ${\bf S}_{i}=C^{\dagger}_{i}{\vec\sigma}
C_{i}/2$ is the spin operator with ${\vec\sigma}=(\sigma_{x},
\sigma_{y},\sigma_{z})$ as the Pauli matrices, $\mu$ is the
chemical potential, and the projection operator $P$ removes zero
occupancy, i.e., $\sum_{\sigma} C^{\dagger}_{i\sigma}C_{i\sigma}
\geq 1$ . The importance of $t'$ term in Eq. (1) has been
emphasized by many authors. It has been argued that although $t'$
term does not change spin configuration because of the same
sublattice hoppings, it can stabilize the AF correlation in the
electron-doped cuprates, and distinguishes between the hole- and
electron-doped cases \cite{tohyama}. For the hole-doped cuprates,
a fermion-spin theory based on the charge-spin separation has been
developed to incorporated a single occupancy local constraint
\cite{feng1}. To apply this theory in the electron-doped cuprates,
the $t$-$t'$-$J$ model (1) can be rewritten in terms of a
particle-hole transformation $C_{i\sigma}\rightarrow
f^{\dagger}_{i-\sigma}$ as,
\begin{eqnarray}
H&=&-t\sum_{i\hat{\eta}\sigma}f_{i\sigma}^{\dagger}
f_{i+\hat{\eta}\sigma}+t'\sum_{i\hat{\tau}\sigma}f_{i\sigma}^{\dagger}
f_{i+\hat{\tau}\sigma} \nonumber \\
&+&\mu\sum_{i\sigma}f_{i\sigma }^{\dagger} f_{i\sigma
}+J\sum_{i\hat{\eta}}{\bf S}_{i} \cdot{\bf S}_{i+\hat{\eta}},
\end{eqnarray}
supplemented by a local constraint $\sum_{\sigma}
f^{\dagger}_{i\sigma}f_{i\sigma}\leq 1$ to remove double
occupancy, where $f^{\dagger}_{i\sigma}$ ($f_{i\sigma}$) is the
hole creation (annihilation) operator, and ${\bf S}_{i}=
f^{\dagger}_{i}{\vec\sigma}f_{i}/2$ is the spin operator in the
hole representation. Now we follow the charge-spin separation
fermion-spin theory \cite{feng1}, and decouple hole operators
$f_{i\uparrow}$ and $f_{i\downarrow}$ as,

\begin{eqnarray}
f_{i\uparrow}=a^{\dagger}_{i\uparrow}S^{-}_{i},~~~~
f_{i\downarrow}=a^{\dagger}_{i\downarrow}S^{+}_{i},
\end{eqnarray}
where the spinful fermion operator
$a_{i\sigma}=e^{-i\Phi_{i\sigma}}a_{i}$ describes the charge
degree of freedom together with some effects of the spin
configuration rearrangements due to the presence of the doped
electron itself (dressed fermion), while the spin operator $S_{i}$
describes the spin degree of freedom (dressed spinon), then the no
double occupancy local constraint, $\sum_{\sigma}
f^{\dagger}_{i\sigma}f_{i\sigma} =S^{+}_{i}a_{i\uparrow}
a^{\dagger}_{i\uparrow} S^{-}_{i}+ S^{-}_{i}a_{i\downarrow}
a^{\dagger}_{i\downarrow} S^{+}_{i}= a_{i}a^{\dagger}_{i}
(S^{+}_{i} S^{-}_{i}+S^{-}_{i} S^{+}_{i})=1- a^{\dagger}_{i}
a_{i}\leq 1$, is satisfied in analytical calculations, and the
double dressed fermion occupancy, $a^{\dagger}_{i\sigma}
a^{\dagger}_{i-\sigma}= e^{i\Phi_{i\sigma}} a^{\dagger}_{i}
a^{\dagger}_{i} e^{i\Phi_{i-\sigma}}=0$ and $a_{i\sigma}
a_{i-\sigma}= e^{-i\Phi_{i\sigma}}a_{i}a_{i} e^{-i\Phi_{i-\sigma}}
=0$, are ruled out automatically. These dressed fermion and spinon
have been shown to be gauge invariant, and in this sense, they are
real and can be interpreted as the physical excitations
\cite{feng1}. We emphasize that this dressed fermion $a_{i\sigma}$
is a spinless fermion $a_{i}$ incorporated a spin cloud
$e^{-i\Phi_{i\sigma}}$ (magnetic flux), and is a magnetic
dressing. In other words, the gauge invariant dressed fermion
carries some spin messages, i.e., it shares its nontrivial spinon
environment \cite{martins}. Although in common sense $a_{i\sigma}$
is not a real spinful fermion, it behaves like a spinful fermion.
In this charge-spin separation fermion-spin representation, the
low-energy behavior of the $t$-$t'$-$J$ model (2) can be expressed
as \cite{feng1},
\begin{eqnarray}
H&=&-t\sum_{i\hat{\eta}}(a_{i\uparrow}S^{+}_{i}
a^{\dagger}_{i+\hat{\eta}\uparrow}S^{-}_{i+\hat{\eta}}+
a_{i\downarrow}S^{-}_{i}a^{\dagger}_{i+\hat{\eta}\downarrow}
S^{+}_{i+\hat{\eta}})\nonumber \\
&+&t'\sum_{i\hat{\tau}}(a_{i\uparrow}S^{+}_{i}
a^{\dagger}_{i+\hat{\tau}\uparrow}S^{-}_{i+\hat{\tau}}+
a_{i\downarrow}S^{-}_{i}a^{\dagger}_{i+\hat{\tau}\downarrow}
S^{+}_{i+\hat{\tau}})\nonumber \\
&-&\mu\sum_{i\sigma}a^{\dagger}_{i\sigma}a_{i\sigma}+J_{{\rm eff}}
\sum_{i\hat{\eta}}{\bf S}_{i}\cdot {\bf S}_{i+\hat{\eta}}¬¬¬¬¬¬,
\end{eqnarray}
with $J_{{\rm eff}}=(1-\delta)^{2}J$, and $\delta=\langle
a^{\dagger}_{i\sigma}a_{i\sigma}\rangle=\langle
a^{\dagger}_{i}a_{i}\rangle$ is the electron doping concentration.
At the half-filling, the $t$-$t'$-$J$ model is reduced to the
Heisenberg model with AFLRO. As we have mentioned above, the phase
with this AFLRO is much more robust in the electron-doped cuprates
and persists to much higher doping levels \cite{alff,tokura1},
therefore in this paper we only discuss the charge dynamics in the
doped regime without AFLRO, i.e., $\langle S_{i}^{z}\rangle =0$.

Since the local constraint has been treated properly within the
charge-spin separation fermion-spin theory, this leads to
disappearing of the extra gauge degree of freedom related to the
local constraint \cite{feng1}, and then the charge fluctuation
only couples to dressed fermions \cite{feng1,feng2}. In this case,
the doping dependence of the charge dynamics of the hole-doped
cuprates has been discussed \cite{feng1,feng2}. Following their
discussions, the optical conductivity of electron-doped cuprates
can be obtained as,
\begin{eqnarray}
\sigma(\omega)&=&{1\over 4}(Ze)^{2}{1\over N}\sum_{k\sigma}
\gamma_{sk}^{2}\int^{\infty}_{-\infty}{d\omega'\over 2\pi}
A^{(a)}_{\sigma}({\bf k},\omega'+\omega) \nonumber \\
&\times& A^{(a)}_{\sigma}({\bf
k},\omega'){n_{F}(\omega'+\omega)-n_{F}(\omega') \over \omega},
\end{eqnarray}
where $Z$ is the number of the nearest neighbor sites,
$\gamma_{sk}^{2}=[(\chi_{1}t-2\chi_{2}t'\cos k_{y})^{2}\sin^{2}
k_{x}+(\chi_{1}t-2\chi_{2}t'\cos k_{x})^{2}\sin^{2} k_{y}]/4$,
$n_{F}(\omega)$ is the fermion distribution function, while the
dressed fermion spectral function $A^{(a)}_{\sigma}(k,\omega)$ is
obtained as $A^{(a)}_{\sigma} (k,\omega)=-2{\rm Im}g_{\sigma}
(k,\omega)$, where the full dressed fermion Green's function
$g^{-1}_{\sigma}(k,\omega)= g^{(0)-1}_{\sigma}(k,\omega)-
\Sigma^{(a)}(k,\omega)$ with the mean-field dressed fermion
Green's function $g^{(0)-1}_{\sigma}(k,\omega)= \omega-\xi_{k}$,
and the second-order dressed fermion self-energy from the dressed
spinon pair bubble \cite{feng1,feng2},
\begin{eqnarray}
&\Sigma&^{(a)}(k,\omega)={1\over 2}Z^{2}{1\over N^{2}}\sum_{pq}
\gamma^{2}_{12}(k,p,q){B_{q+p}B_{q}\over 4\omega_{q+p}\omega_{q}}
\nonumber \\
&\times&\left({F^{(1)}(k,p,q)\over \omega+\omega_{q+p}-
\omega_{q}-\xi_{p+k}}+{F^{(2)}(k,p,q)\over \omega-\omega_{q+p}+
\omega_{q}-\xi_{p+k}}\right. \nonumber \\
&+&\left.{F^{(3)}(k,p,q)\over \omega+\omega_{q+p}+\omega_{q}
-\xi_{p+k}}+{F^{(4)}(k,p,q)\over \omega-\omega_{q+p}-\omega_{q}-
\xi_{p+k}}\right ),
\end{eqnarray}
where $\gamma^{2}_{12}(k,p,q)=[(t\gamma_{q+p+k}-t'\gamma'_{
q+p+k})^{2}+(t\gamma_{q-k}-t'\gamma'_{q-k})^{2}]$, $B_{k}=
\lambda_{1}[2\chi^{z}_{1}(\epsilon\gamma_{ k}-1)+ \chi_{1}
(\gamma_{k}-\epsilon)]- \lambda_{2}(2\chi^{z}_{2} \gamma'_{k}
-\chi_{2})$, $\lambda_{1}=2ZJ_{eff}$, $\lambda_{2}=4Z\phi_{2}t'$,
$\gamma_{ k}=(1/Z) \sum_{\hat{\eta}}e^{i{\bf k}\cdot \hat{\eta}}$,
$\gamma'_{ k} =(1/Z)\sum_{\hat{\tau}} e^{i{\bf k}\cdot\hat{\tau}}
$, $F^{(1)} (k,p,q)=n_{F}(\xi_{p+k})[n_{B}(\omega_{q})-n_{B}
(\omega_{q+p})]+n_{B}(\omega_{q+p})[1+n_{B}(\omega_{q})]$,
$F^{(2)}(k,p,q)=n_{F}(\xi_{p+k})[n_{B}(\omega_{q+p})-n_{B}
(\omega_{q})]+n_{B}(\omega_{q})[1+n_{B}(\omega_{q+p})]$,
$F^{(3)}(k,p,q)=n_{F}(\xi_{p+k})[1+n_{B}(\omega_{q+p})+n_{B}
(\omega_{q})]+n_{B}(\omega_{q})n_{B}(\omega_{q+p})$, $F^{(4)}
(k,p,q)=[1+n_{B}(\omega_{q})][1+n_{B}(\omega_{q+p})] -n_{F}
(\xi_{p+k})[1+n_{B}(\omega_{q+p})+n_{B}(\omega_{q})]$,
$n_{B}(\omega_{k})$ is the Bose distribution function, and the
mean-field dressed fermion and spinon excitation spectra are given
by, $\xi_{k}=Zt\chi_{1}\gamma_{{\bf k}}-Zt'\chi_{2} \gamma'_{{\bf
k}}-\mu,\omega^{2}_{k}=A_{1}(\gamma_{k})^{2}+A_{2}(\gamma'_{k})^{2}+
A_{3}\gamma_{k}\gamma'_{k}+A_{4}\gamma_{k}+A_{5}\gamma'_{k}+A_{6}
$ respectively, with $A_{1}=\alpha\epsilon\lambda_{1}^{2}(\epsilon
\chi^{z}_{1}+\chi_{1}/2)$, $A_{2}=\alpha\lambda_{2}^{2}
\chi^{z}_{2}$, $A_{3}=-\alpha\lambda_{1}\lambda_{2}(\epsilon
\chi^{z}_{1}+\epsilon\chi^{z}_{2}+\chi_{1}/2)$, $A_{4}=-\epsilon
\lambda_{1}^{2}[\alpha(\chi^{z}_{1}+\epsilon\chi_{1}/2)+(\alpha
C^{z}_{1}+(1-\alpha)/(4Z)-\alpha\epsilon\chi_{1}/(2Z))+(\alpha
C_{1}+(1-\alpha)/(2Z)-\alpha\chi^{z}_{1}/2)/2]+\alpha\lambda_{1}
\lambda_{2}(C_{3}+\epsilon\chi_{2})/2$, $A_{5}=-3\alpha
\lambda^{2}_{2}\chi_{2}/(2Z)+\alpha\lambda_{1}\lambda_{2}
(\chi^{z}_{1}+\epsilon\chi_{1}/2+C^{z}_{3})$, $A_{6}=
\lambda^{2}_{1}[\alpha C^{z}_{1}+(1-\alpha)/(4Z)-\alpha\epsilon
\chi_{1}/(2Z)+\epsilon^{2}(\alpha C_{1}+(1-\alpha)/(2Z)-\alpha
\chi^{z}_{1}/2)/2]+\lambda^{2}_{2}(\alpha C_{2}+(1-\alpha)/(2Z)-
\alpha\chi^{z}_{2}/2)/2)-\alpha\epsilon\lambda_{1}\lambda_{2}
C_{3}$, and the spinon correlation functions $\chi^{z}_{1}=\langle
S_{i}^{z}S_{i+\hat{\eta}}^{z}\rangle$, $\chi^{z}_{2}=\langle
S_{i}^{z}S_{i+\hat{\tau}}^{z}\rangle$,
$C_{1}=(1/Z^{2})\sum_{\hat{\eta},\hat{\eta'}}\langle
S_{i+\hat{\eta}}^{+}S_{i+\hat{\eta'}}^{-}\rangle$,
$C^{z}_{1}=(1/Z^{2})\sum_{\hat{\eta},\hat{\eta'}}\langle
S_{i+\hat{\eta}}^{z}S_{i+\hat{\eta'}}^{z}\rangle$,
$C_{2}=(1/Z^{2})\sum_{\hat{\tau},\hat{\tau'}}\langle
S_{i+\hat{\tau}}^{+}S_{i+\hat{\tau'}}^{-}\rangle$,
$C_{3}=(1/Z)\sum_{\hat{\tau}}\langle S_{i+\hat{\eta}}^{+}
S_{i+\hat{\tau}}^{-}\rangle$, $C^{z}_{3}=(1/Z)
\sum_{\hat{\tau}}\langle S_{i+\hat{\eta}}^{z}
S_{i+\hat{\tau}}^{z}\rangle$. In order to satisfy the sum rule for
the correlation function $\langle S_{i}^{+}S_{i}^{-}\rangle=1/2$
in the absence of AFLRO, a decoupling parameter $\alpha$ has been
introduced in the MF calculation, which can be regarded as the
vertex correction \cite{feng3,kondo}. All these mean-field order
parameters, decoupling parameter, and the chemical potential are
determined self-consistently \cite{feng3}.
\begin{figure}[prb]
\epsfxsize=3.0in\centerline{\epsffile{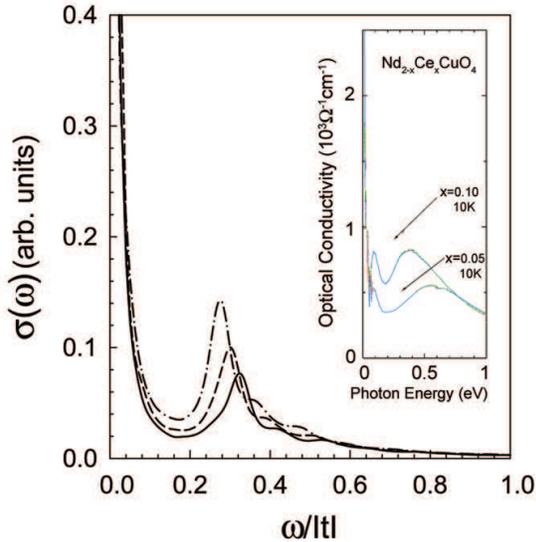}} \caption{The
conductivity of electron-doped cuprates at $\delta=0.10$ (solid
line), $\delta=0.12$ (dashed line), and $\delta=0.15$ (dash-dotted
line) for $t/J=-2.5$ and $t'/t=0.15$ with $T=0$. Inset: the
experimental result on Nd$_{2-x}$Ce$_{x}$CuO$_{4}$ taken from Ref.
\cite{onose}.}
\end{figure}

\begin{figure}[prb]
\epsfxsize=3.0in\centerline{\epsffile{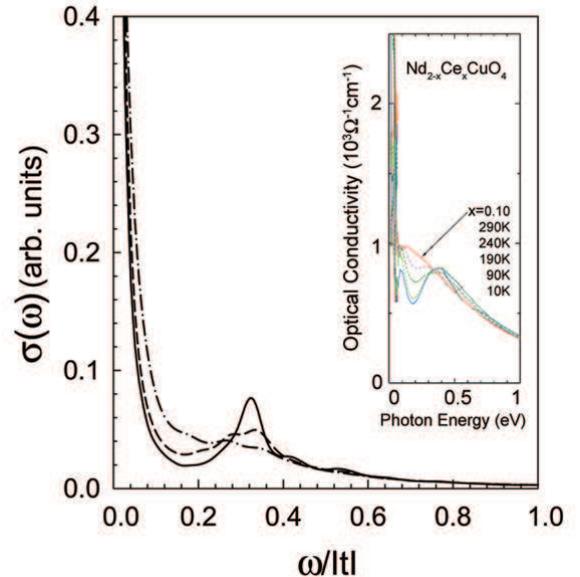}} \caption{The
conductivity of electron-doped cuprates at $\delta=0.10$ for
$t/J=-2.5$ and $t'/t=0.15$ with $T=0$ (solid line), $T=0.1J$
(dashed line), and $T=0.2J$ (dash-dotted line). Inset: the
experimental result on Nd$_{2-x}$Ce$_{x}$CuO$_{4}$ taken from Ref.
\cite{onose}.}
\end{figure}
The optical conductivity measurement is a powerful probe for
interacting electron systems \cite{tanner}, and can provide very
detailed knowledges about the low-energy excitations as electrons
are doped to Mott insulators. We have performed a numerical
calculation for the optical conductivity in Eq. (5), and the
results at electron doping $\delta=0.10$ (solid line),
$\delta=0.12$ (dashed line), and $\delta=0.15$ (dash-dotted line)
for parameters $t/J=-2.5$ and $t'/t=0.15$ at temperature $T=0$ are
shown in Fig. 1, where the charge $e$ has been set as the unit.
For a comparison, the experimental result \cite{onose} of
Nd$_{2-x}$Ce$_{x}$CuO$_{4}$ is also plotted in Fig. 1 (inset).
This conductivity spectrum shows a pseudogap structure, where a
low-energy peak appears at $\omega\sim 0$, which decays rapidly,
and a rather sharp midinfrared peak appears around $\omega\sim
0.3\mid t\mid$, which is in contrast to the case for the
hole-doped cuprates, where a broad distribution of the spectral
weight of the midinfrared band is observed \cite{feng1,feng2}.
Since the spectral function (then full dressed fermion Green's
function) in the optical conductivity (5) is obtained by
considering the second-order corrections due to the dressed spinon
pair bubble, then our present results reflect that the AF spin
correlation in the electron-doped case is stronger than that in
the hole-doped case, and the pseudogap structure or very sharp
midinfrared peak is closely related to the strong AF spin
correlations. This is consistent with the results from numerical
simulations \cite{tohyama1}, where the pseudogap structure in the
electron-doped cuprates is very sensitive to not only $J$, but
also $t'$. Although $t'$ does not change spin configuration, it
can enhance the AF correlation in the electron-doped cuprates.
With increasing the value of $t'/t$, the gap increases in energy
\cite{tohyama1}. Moreover, the unusual midinfrared peak is
electron doping dependent, and the component increases with
increasing electron doping for $0.08\mid t\mid<\omega <0.5 \mid
t\mid$, and is nearly independent of electron doping for $\omega
>0.5\mid t\mid$. This reflects an increase in the mobile carrier
density, and indicates that the spectral weight of the midinfrared
sidepeak is taken from the Drude absorption, then the spectral
weight from both low energy peak and midinfrared peak represent
the actual free-carrier density. For a better understanding of the
optical properties of the electron-doped cuprates, we have studied
the conductivity at different temperatures, and the results at
$\delta=0.10$ for $t/J=-2.5$ and $t'/t=0.15$ with $T=0$ (solid
line), $T=0.1J$ (dashed line), and $T=0.2J$ (dash-dotted line) are
plotted in Fig. 2 in comparison with the experimental data
\cite{onose} taken from Nd$_{2-x}$Ce$_{x}$CuO$_{4}$ (inset). Our
results show that the conductivity spectrum is temperature
dependent for $\omega <0.5\mid t\mid$, and almost temperature
independent for $\omega>0.5\mid t\mid$. The component in the low
energy region increases with increasing temperatures, then there
is a tendency towards the Drude-like behavior, while the unusual
midinfrared peak is severely suppressed with increasing
temperatures, and vanishes at high temperatures. This reflects
that the spin correlation rapidly decreases with increasing
temperature. These results are in qualitative agreement with
experiments \cite{onose} and the numerical simulations
\cite{tohyama1}. As in the hole-doped cuprates, the charge
transport is governed by the dressed fermion scattering, therefore
$\delta$ dressed fermions are responsible for the optical
conductivity, i.e., the optical conductivity in the electron-doped
cuprates is carried by $\delta$ electrons. Since the strong
electron correlation is common for both hole- and electron-doped
cuprates, these similar behaviors observed from the optical
conductivity are expected.

\begin{figure}[prb]
\epsfxsize=3.0in\centerline{\epsffile{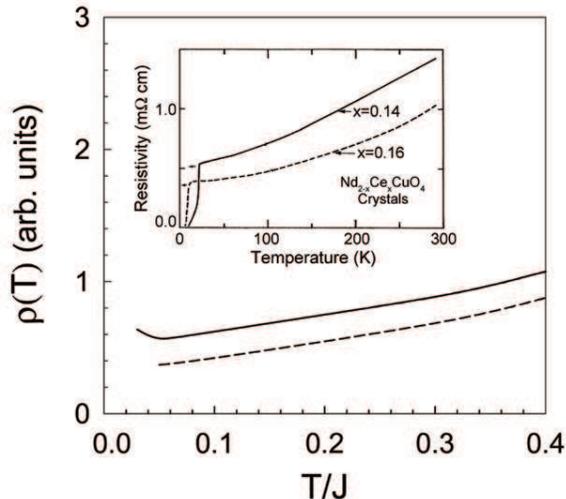}} \caption{The
resistivity of electron-doped cuprates for $t/J=-2.5$ and
$t'/t=0.15$ at $\delta=0.09$ (solid line) and $\delta=0.15$
(dashed line). Inset: the experimental result on
Nd$_{2-x}$Ce$_{x}$CuO$_{4}$ taken from Ref. \cite{wang}.}
\end{figure}

The quantity which is closely related to the conductivity is the
resistivity, which can be obtained as
$\rho(T)=1/\lim_{\omega\rightarrow 0}\sigma (\omega)$. This
resistivity has been calculated numerically, and the results for
$t/J=-2.5$ and $t'/t=0.15$ at $\delta=0.09$ (solid line) and and
$\delta=0.15$ (dashed line) are plotted in Fig. 3, in comparison
with the experimental data \cite{wang} taken from
Nd$_{2-x}$Ce$_{x}$CuO$_{4}$ (inset). It is shown obviously that
the resistivity is characterized by a crossover from the high
temperature metallic-like to low temperature insulating-like
behavior in the relatively low doped regime, and a metallic-like
behavior in the relatively high doped regime. But even in the
relatively low doped regime, the resistivity exhibits the
metallic-like behavior over a wide range of temperatures, which
also is in qualitative agreement with experiments
\cite{onose,wang}.

The physical interpretation to the above obtained results is very
similar to these in the hole-doped case \cite{feng1,feng2}, since
the $t$-$t'$-$J$ model in both hole- and electron-doped cuprates
is characterized by a competition between the kinetic energy and
magnetic energy, with the magnetic energy favors the magnetic
order for spins, while the kinetic energy favors delocalization of
electrons and tends to destroy the magnetic order. However, it
needs more kinetic energy to drive the motion of charged carriers
in the electron-doped case since the strong spin correlation. In
the charge-spin separation fermion-spin theory, although both
dressed fermions and spinons contribute to the charge dynamics,
the dressed fermion scattering dominates the charge dynamics
\cite{feng1,feng2}, where the dressed fermion scattering rate is
obtained from the full dressed fermion Green's function (then the
dressed fermion self-energy and spectral function) by considering
the dressed fermion-spinon interaction. In this case, the
crossover from the high temperature metallic-like to low
temperature insulating-like behaviors in the relatively low
electron-doped regime, and the metallic-like behavior in the
relatively high electron-doped regime in the resistivity is
closely related to this competition between the kinetic energy and
magnetic energy. In lower temperatures, the dressed fermion
kinetic energy is much smaller than the magnetic energy in the
relatively low electron-doped regime, then the magnetic
fluctuation is strong enough to severely reduce the dressed
fermion scattering and thus is responsible for the insulating-like
behavior in the resistivity. With increasing temperatures, the
dressed fermion kinetic energy is increased, while the dressed
spinon magnetic energy is decreased. In the region where the
dressed fermion kinetic energy is much larger than the dressed
spinon magnetic energy at higher temperatures or relatively high
electron-doped regime, the dressed fermion scattering would give
rise to the metallic-like resistivity.

In summary, we have studied the doping dependence of the charge
dynamics in the electron-doped cuprates within the $t$-$t'$-$J$
model. The optical conductivity spectrum shows an unusual
pseudogap structure with a low-energy peak appearing at
$\omega\sim 0$ and an rather sharp midinfrared peak appearing
around $\omega\sim 0.3\mid t\mid$, and the resistivity exhibits a
crossover from the high temperature metallic-like to low
temperature insulating-like behavior in the relatively low
electron-doped regime, and a metallic-like behavior in the
relatively high electron-doped regime. Our results also show that
the unusual pseudogap structure is intriguingly related to the
strong AF correlation in the system.

\acknowledgments

The authors would like to thank Dr. Ying Liang, Dr. Bin Liu, and
Dr. Jihong Qin for the helpful discussions. This work was
supported by the National Natural Science Foundation of China
under Grant Nos. 10125415 and 90403005, and the Grant from Beijing
Normal University.

\end{document}